\documentclass[aip,jap,preprint,groupedaddress,showpacs,superscriptaddress,citeautoscript]{revtex4-1}
\usepackage{amsmath}
\usepackage[pdftex]{graphicx}
\usepackage{epstopdf}

\usepackage[colorlinks=true,linkcolor=blue,citecolor=blue,urlcolor=black]{hyperref}%
\expandafter\ifx\csname package@font\endcsname\relax\else
 \expandafter\expandafter
 \expandafter\usepackage
 \expandafter\expandafter
 \expandafter{\csname package@font\endcsname}%
\fi

\newcommand{\Eq} [1] {Eq.~\ref{#1}}
\newcommand{\Fig}[1] {Fig.~\ref{#1}}
\newcommand{\Sec}[1] {Sec.~\ref{#1}}
\begin{document}
\title{Strain induced mobility modulation in single-layer MoS$_{2}$}

\author{Manouchehr Hosseini}
\affiliation{School of Electrical and Computer Engineering, University of Tehran, P.O.Box 14395-515, Tehran, Iran}

\author{Mohammad Elahi}
\affiliation{School of Electrical and Computer Engineering, University of Tehran, P.O.Box 14395-515, Tehran, Iran}

\author{Mahdi Pourfath}
\email{pourfath@iue.tuwien.ac.at}
\affiliation{School of Electrical and Computer Engineering, University of Tehran, P.O.Box 14395-515, Tehran, Iran}
\affiliation{Institute for Microelectronics, TU Wien, Gusshausstrasse 27--29/E360, 1040 Vienna, Austria}

\author{David Esseni}
\email{david.esseni@uniud.it}
\affiliation{DIEGM, Via delle Scienze 206, 33100 Udine, Italy}

\date{\today}

\begin{abstract} 
In this paper the effect of biaxial and uniaxial strain on the mobility of single-layer MoS$_{2}$ for temperatures T $>$ 100 K is investigated. Scattering from intrinsic phonon modes, remote phonon and charged impurities are considered along with static screening. Ab-initio simulations are utilized to investigate the strain induced effects on the electronic bandstructure and the linearized Boltzmann transport equation is used to evaluate the low-field mobility under various strain conditions. The results indicate that the mobility increases with tensile biaxial and tensile uniaxial strain along the armchair direction. Under compressive strain, however, the mobility exhibits a non-monotonic behavior when the strain magnitude is varied. In particular, with a relatively small compressive strain of 1\% the mobility is reduced by about a factor of two compared to the unstrained condition, but with a larger compressive strain the mobility partly recovers such a degradation.
\end{abstract}

\maketitle

\section{Introduction}
Single and few-layers of transition metal dichalcogenides show promising electronic, optical, and mechanical properties and are considered as potential candidates for future electronic applications \cite{neto2011new}. Because of weak inter-layer van der Waals bonds in their layered structure, single to few-layers of these materials can be obtained by mechanical or chemical exfoliation techniques \cite{novoselov2005two,ayari2007realization,ramakrishna2010mos2}. Among these materials single-layer MoS$_{2}$ has attracted the attention of  scientists\cite{korn2011low,splendiani2010emerging,mak2010atomically,mak2013tightly}. Single-layer MoS$_{2}$ has a direct band gap of 1.8 to 1.9 eV \cite{mak2010atomically,splendiani2010emerging}, which makes it suitable for various electronic applications \cite{yoon2011good}. It has been shown that the application of compressive and tensile biaxial strain results in an indirect bandgap \cite{tabatabaei2013first,feng2012strain,ghorbani2013strain}. An MOS transistor based on this material has demonstrated a $I_\mathrm{on}/I_\mathrm{off}$ ratio of $∼10^{8}$, a relatively steep sub-threshold swing of 74 mV/dec and an extremely small off-current of 25 fA/$\mu$m \cite{radisavljevic2011single}, moreover possible applications to hetero-junction inter-layer tunneling FETs have also been proposed and theoretically investigated \cite{li2014single}. Room temperature mobility of n-type single-layer MoS$_{2}$ has been reported to be in the range of 0.5--3 cm$^{2}$/(Vs) and can be increased to about 200 cm$^{2}$/(Vs) with the use of high-$\kappa$ dielectrics \cite{radisavljevic2011single}.

The low-field mobility is one of the most important transport properties for a large number of physical systems and electronic devices. A comprehensive study of strain effects on the mobility of single layer MoS$_{2}$, however, is missing. In the present work, the effects of biaxial and uniaxial strain on the low-field mobility of single-layer MoS$_{2}$ is investigated by using ab-initio simulations along with the linearized Boltzmann transport equation (BTE) \cite{paussa2013exact}. Scattering rates due to intrinsic phonon, charge impurities, and remote phonon are taken into account. 
\section{Bandstructure and Scattering Rates}
\label{s:APP}
In the first part of this section, some details about the ab-initio calculations of the electronic bandstructure in the presence of strain are discussed. Thereafter, the formulation of various scattering rates is described.
\subsection{Bandstructure}
\label{s:App:BAND1}
We carry out first-principle simulations based on the density-functional theory (DFT) along with the local density approximation (LDA) as implemented in the SIESTA code \cite{soler2002siesta,han2011band,lebegue2009electronic} to investigate the relevant electronic properties of a single layer MoS$_2$ under strain. While DFT-LDA, in general, underestimates band gaps, the resulting dispersion of individual bands, i.e., effective masses and energy differences between valleys, is less problematic \cite{kaasbjerg2012phonon}. A cutoff energy equal to $600$ Ry is used and a vacuum separation of $30$ \AA{} is adopted, which is sufficient to hinder interactions between adjacent layers. Sampling of the reciprocal space Brillouin zone (BZ) is performed by a Monkhorst-Pack grid of $\mathrm{18\times18\times1}$ $k$-points. Calculations begin with the determination of the optimized geometry, that is the configuration in which the residual Hellmann-Feynman forces acting on atoms are smaller than $0.01$ eV/\AA{}. The calculated lattice constant of unstrained single-layer MoS$_{2}$ is 3.11 \AA{} that has a good agreement with the reported value in Ref.~\onlinecite{chang2013orbital,ataca2011functionalization}. \Fig{f:Ek0}(a) shows the energy contours of the conduction-band in the first BZ for unstrained single layer MoS$_{2}$. In an unstrained material the lowest and the second lowest minimum in the conduction band are denoted as K-valley and Q-valley, respectively. The 6 K valleys are degenerate in the unstrained and in all the strained conditions explored in this paper. The 6 Q-valleys are degenerate in unstrained conditions, while under uniaxial strain they split into 4 Q$_\mathrm{A}$-valleys and 2 Q$_\mathrm{B}$-valleys with different effective masses and energy minima, as discussed and illustrated in~\Sec{s:RST}. The energy distance between K-valley and Q-valley for unstrained material is evaluated to be 160 meV, in agreement with Ref. \onlinecite{kaasbjerg2012phonon}. \Fig{f:Ek0}(b) shows the calculated DFT-LDA band structure and depicts a direct band gap of 1.92 eV at the K-point which is very close to the experimentally measured value of 1.85 eV \cite{splendiani2010emerging}.
\subsection{Scattering with MoS$_{2}$ phonon modes}
\label{s:App:SCA}
Scattering rates due to intrinsic phonons (including acoustic, optical and polar-optical phonons), to remote phonons and to charged impurities are taken into account. Piezoelectric coupling to the acoustic phonons is only important at low temperatures and is neglected in this work \cite{kaasbjerg2013acoustic}. If the surrounding dielectric provides a large energy barrier for confining electrons in the MoS$_{2}$ layer, the envelope function of mobile electrons can be approximated as $\Psi_{\vec{k}}(\vec{r},z) = \chi(z) \exp{\left(i\vec{k}.\vec{r}\right)}/\sqrt{S}$ with $\chi(z) =\sqrt{(2/a)}\sin(\pi z/a)$ \cite{ma2014charge}, where $S$ is the area normalization factor, $\vec{k}$ is the in-plane two-dimensional wave vector, $a$ is the thickness of single layer MoS$_{2}$ and $\vec{r}$ is the in-plane position vector. The scattering rates for the acoustic and optical phonon are discussed first. 

Using Fermi's golden rule the scattering rate from an initial state $\vec{k}$ in valley $v$ to the final state $\vec{k'}$ in valley $w$ can be written as
\begin{equation}
S^{v,w}(\vec{k},\vec{k'}) =\frac{2\pi}{\hbar}|M^{v,w}(\vec{k},\vec{k'})|^{2} \delta[E^{w}(\vec{k'}) - E^{v}(\vec{k}) \mp \hbar\omega(q)] \ ,
  \label{e:Eq16}
\end{equation}
where $|M^{v,w}(\vec{k},\vec{k'})|$ is the matrix element for the mentioned transition and $\hbar\omega(q)$ is the phonon energy that may depend on $q = |\vec{k} - \vec{k'}|$. The intra-valley transitions ($v=w$) assisted by acoustic phonons can be approximated as elastic and the rate is given by
\begin{equation}
S_\mathrm{ac}(\vec{k},\vec{k}') = \frac{2\pi k_\mathrm{B} T D_\mathrm{ac}^{2}}{\rho S \hbar v_{s}^{2}} \delta[E(\vec{k}') - E(\vec{k})]\ ,
\label{e:Eq01}
\end{equation}
where $k_\mathrm{B}$ is the Boltzmann constant, $T$ is the absolute temperature, $D_\mathrm{ac}$ is the acoustic the deformation potential, $\rho = 3.1 \times 10^{-7}$ [gr/cm$^{2}$] is the mass density and $v_{s}$ is the sound velocity of single layer MoS$_{2}$. On the other hand, the rate of inelastic phonon scattering, including intra and inter-valley optical phonons, and inter-valley acoustic phonons, can be expressed as
\begin{equation} 
S^{v,w}_{\mathrm{ac/op}}(\vec{k},\vec{k'}) = \frac{\pi (D^{v,w}_{\mathrm{ac/op}})^{2}}{\omega_\mathrm{ac/op} \rho S }\left[ n_\mathrm{op} + \frac{1}{2} \mp \frac{1}{2} \right] \delta[E^{w}(\vec{k}') - E^{v}(\vec{k}) \mp \hbar \omega_\mathrm{ac/op}(q)]\ ,
\label{e:Eq02}
\end{equation}
where $D^{v,w}_{\mathrm{ac/op}}$ is the acoustic/optical deformation potential for a transition between valleys $v$ and $w$, $\hbar\omega_\mathrm{ac/op}(q)$ is the phonon energy, and $n_\mathrm{op}$ is the phonon occupation (upper and lower sign denote phonon absorption and phonon emission, respectively). The phonon assisted inter-valley transitions considered in this work, and the corresponding phonon momentum are shown in \Fig{f:DP}. In our calculations, we employed the deformation potentials and phonon energies from Ref.~\onlinecite{li2013intrinsic} that are reported in Table~\ref{t:DP} and Table~\ref{t:PH}. It should be noted that the same deformation potentials are used for Q$_\mathrm{A}$ and Q$_\mathrm{B}$ valleys. 
\subsection{Remote Phonon Scattering}
\label{s:App:SCA2}
Another important scattering source considered in this work is the remote phonon or surface-optical (SO) phonon scattering mechanism. The source of this scattering is in the surrounding dielectrics via long-range Coulomb interactions, provided that the dielectrics support polar vibrational modes. By assuming semi-infinite oxides and neglecting the possible coupling to the plasmons of the two-dimensional material, the energy dispersion of SO phonons can be obtained by solving the secular equation \cite{ong2013theory}
\begin{equation}
(\epsilon_\mathrm{box}(\omega) + \epsilon_\mathrm{2D})(\epsilon_\mathrm{tox}(\omega) + \epsilon_\mathrm{2D}) - (\epsilon_\mathrm{box}(\omega) - \epsilon_\mathrm{2D})(\epsilon_\mathrm{tox}(\omega) - \epsilon_\mathrm{2D})e^{-2qa}= 0 \ ,
  \label{e:Eq03}
\end{equation}
where the thickness $a$ of the single layer MoS$_{2}$ is set to $a = 3.17$ \AA{} \cite{yue2012mechanical}, $\epsilon_\mathrm{2D}$ is the dielectric constant of the two-dimensional material (single layer MoS$_{2}$ in this work), the index box and tox denote the back-oxide ($z < 0$) and the top-oxide ($z> a$), respectively, $q = |\vec{k} - \vec{k'}|$ and $\epsilon_\mathrm{2D}$ for MoS$_{2}$ is set to 7.6 \cite{ma2014charge}. A numerical solution of \Eq{e:Eq03} shows that the frequency of remote phonon has a very weak dependence on $q$, that consequently we neglected in our calculations by setting $e^{-2qa} \approx 1$ in \Eq{e:Eq03}. With this approximation, \Eq{e:Eq03} simplifies to $\epsilon_\mathrm{box}(\omega) + \epsilon_\mathrm{tox}(\omega) = 0$, that we solved by using the single polar phonon expression for the $\epsilon_\mathrm{ox}(\omega)$ in each oxide: 
\begin{equation}
\epsilon_\mathrm{ox} (\omega) = \epsilon^{\infty} + \frac{\epsilon^{0} - \epsilon^{\infty}}{1 - \omega^{2}/\omega_\mathrm{TO}^{2}} \ ,
  \label{e:Eq04}
\end{equation}
where $\epsilon^{\infty}$ and $\epsilon^{0}$ are the high and low frequency dielectric constant, respectively, and $\omega_\mathrm{TO}$ is the frequency of the polar phonon in the oxide. We could provide analytical solution for \Eq{e:Eq04} and express $\omega_\mathrm{so,box}$ as: $\omega_\mathrm{so,box}^{2} = (-B + \sqrt{B^2 - 4 A C})/(2A)$ and for $\omega_\mathrm{so,tox}$ as $\omega_\mathrm{so,tox}^{2} = (-B - \sqrt{B^2 - 4 A C})/(2A)$, where $A = (\epsilon_\mathrm{tox}^{\infty} + \epsilon_\mathrm{box}^{\infty})$, $B = -(\epsilon_\mathrm{tox}^{0} + \epsilon_\mathrm{box}^{\infty}) \omega_\mathrm{TO,tox}^{2} - (\epsilon_\mathrm{box}^{0} + \epsilon_\mathrm{tox}^{\infty}) \omega_\mathrm{TO,box}^{2}$ and $C = (\epsilon_\mathrm{tox}^{0} + \epsilon_\mathrm{box}^{0}) \omega_\mathrm{TO,tox}^{2}  \omega_\mathrm{TO,box}^{2}$.  Table~\ref{t:Remote} reports the parameters of dielectric materials that are studied in this work and indicates the corresponding calculated SO phonon frequencies. The scattering matrix element of remote phonon can be written as \cite{ong2013theory}:
\begin{equation}
M_\mathrm{so,tox}(\vec{k},\vec{k'}) = \sqrt{\frac{\hbar \omega_\mathrm{so,tox}}{2Sq} \left(\frac{1}{\epsilon_\mathrm{tox}^{\infty} + \epsilon_\mathrm{box}(\omega_\mathrm{so,tox})} - \frac{1}{\epsilon_\mathrm{tox}^{0} + \epsilon_\mathrm{box}(\omega_\mathrm{so,tox})}\right)} \ ,
  \label{e:Eq06}
\end{equation}
\begin{equation}
M_\mathrm{so,box}(\vec{k},\vec{k'}) = \sqrt{\frac{\hbar \omega_\mathrm{so,box}}{2Sq} \left(\frac{1}{\epsilon_\mathrm{box}^{\infty} + \epsilon_\mathrm{tox}(\omega_\mathrm{so,box})} - \frac{1}{\epsilon_\mathrm{box}^{0} + \epsilon_\mathrm{tox}(\omega_\mathrm{so,box})}\right)} \ ,
  \label{e:Eq061}
\end{equation}
Scattering with SO phonon mode is inelastic and we consider only intra-valley transitions. The corresponding transition rate is 
\begin{equation}
S_{\mathrm{so}}(\vec{k},\vec{k}') = \frac{2\pi}{\hbar} |M_\mathrm{so}(\vec{k},\vec{k}')|^{2}\left[ n_\mathrm{so} + \frac{1}{2} \mp \frac{1}{2}\right] \delta[E(\vec{k}') - E(\vec{k}) \mp \hbar \omega_\mathrm{so}] \ ,
  \label{e:Eq062}
\end{equation}
where $n_\mathrm{so}$ and $\hbar\omega_\mathrm{so}$ are the SO phonon occupation number and energy, respectively.
\subsection{Scattering with Coulomb Centers}
\label{s:App:SCA3}
To investigate the effect of the dielectric environment on the scattering of carriers from charged impurities located inside the single layer MoS$_{2}$, we assume that the charged impurities are located in the center of the single layer MoS$_{2}$ thickness, that is at $z=a/2$. The Fourier transform of the scattering potential due to a charged impurity located at $(\vec{r},z) = (0,a/2)$ can be written as \cite{esseni2011nanoscale}
\begin{equation}
\phi(q,z) = \frac{\mathrm{e}^{2}}{2q\epsilon_\mathrm{2D}}\left[e^{-q|z - a/2|} + C e^{qz} + D e^{-qz}\right] \ ,
\label{e:Eq07}
\end{equation}
where ${\mathrm{e}}$ is the elementary charge, and $C$ and $D$ can be written as
\begin{equation}
C = \frac{(\epsilon_\mathrm{2D}-\epsilon_\mathrm{box}^{0})(\epsilon_\mathrm{2D}-\epsilon_\mathrm{tox}^{0})e^{-qa/2} + (\epsilon_\mathrm{2D} + \epsilon_\mathrm{box}^{0})(\epsilon_\mathrm{2D} - \epsilon_\mathrm{tox}^{0})e^{aq/2} - (\epsilon_\mathrm{box}^{0} - \epsilon_\mathrm{2D})(\epsilon_\mathrm{tox}^{0} + \epsilon_\mathrm{2D})}{(\epsilon_\mathrm{2D} + \epsilon_\mathrm{box}^{0})(\epsilon_\mathrm{2D} + \epsilon_\mathrm{tox}^{0})e^{2aq}  - (\epsilon_\mathrm{2D} - \epsilon_\mathrm{box}^{0})(\epsilon_\mathrm{2D} - \epsilon_\mathrm{tox}^{0})} \ ,
  \label{e:Eq08}
\end{equation}
and 
\begin{equation}
D = \frac{(\epsilon_\mathrm{box}^{0} - \epsilon_\mathrm{2D})\left[C + e^{-qa/2} \right]}{\epsilon_\mathrm{box}^{0} + \epsilon_\mathrm{2D}} \ .
  \label{e:Eq09}
\end{equation}
Using \Eq{e:Eq07}, the $\chi(z)$ form $\chi(z) =\sqrt{(2/a)}\sin(\pi z/a)$ and assuming intra-valley transitions for scattering with charged impurities, the transition matrix elements take the form
\begin{equation}
\begin{split}
M_\mathrm{cb}^{(0)}(\vec{k},\vec{k}') = & \frac{e^{2}}{qa\epsilon_\mathrm{2D}}\left(\frac{1}{q} - \frac{q}{q^{2} + \left(2\pi/a \right)^2}\right) \times \left[\frac{C}{2} \left(e^{qa}-1 \right) + \frac{D}{2} \left(1-e^{-qa} \right) - e^{-qa/2} \right] \\ 
&+ \frac{e^{2}}{qa\epsilon_\mathrm{2D}}\left(\frac{1}{q} + \frac{q}{q^{2} + \left(2\pi /a \right)^2}\right)  \ ,
\end{split}
  \label{e:Eq10}
\end{equation}
where $q = |\vec{k} - \vec{k'}|$. \Eq{e:Eq10} expresses the matrix element for a Coulomb center located in $(\vec{r},z) = (0,a/2)$ and does not account for the screening produced by the free carriers in MoS$_{2}$; such a screening effect is introduced according to the dielectric matrix approach discussed in \Sec{s:App:SCA4}. The overall matrix element produced by a set of Coulomb centers randomly distributed at positions $(\vec{r},a/2)$ is known to be affected by the statistical properties of the distribution and, in particular, by a possible correlation between the position of Coulomb centers. In this paper we do not address these difficulties and simply write the overall matrix element as $|M_\mathrm{cb}(\vec{k},\vec{k}')|^2 = \left[N_{D}|M_\mathrm{cb}^{(0)}(\vec{k},\vec{k}')|^{2}\right]/S$, where $N_{D}$ is the impurity density per unit area and $M_\mathrm{cb}^{(0)}$ is given by \Eq{e:Eq10}. Scattering charged impurities is treated as elastic and the rate is therefore given by
\begin{equation}
S_{\mathrm{cb}}(\vec{k},\vec{k}') = \frac{2\pi}{\hbar} |M_\mathrm{cb}(\vec{k},\vec{k}')|^{2} \delta({E(\vec{k}') - E(\vec{k})}) \ .
\end{equation}
\subsection{Screening}
\label{s:App:SCA4}
The effect of static screening produced by the electrons in the MoS$_{2}$ conduction band is described by using the dielectric function approach \cite{esseni2011nanoscale}, so that the screened matrix element $M^{w}_\mathrm{scr}(\vec{k},\vec{k}')$ in valley $w$ is obtained by solving the linear problem:
\begin{equation}
M^{v}(q) = \sum_{w} \epsilon^{v,w}(q) M^{w}_\mathrm{scr}(q)  \ ,
  \label{e:Eq11}
\end{equation}
where $M^{v}(q)$ are the unscreened matrix element. As can be seen in \Fig{f:Ek0}(a), there are three different valleys in strained single layer MoS$_{2}$ (K, Q${_A}$ and Q${_B}$ valleys), hence $v$, $w$ $\in$ \{K, Q$_\mathrm{A}$, Q$_\mathrm{B}$\}. In \Eq{e:Eq11}, $\epsilon^{v,w}$ is the dielectric matrix which is introduced as:
\begin{equation}
\epsilon^{v,w}(q) = \delta_{v,w} - \frac{e^{2}}{q(\epsilon_\mathrm{2D} + \epsilon_\mathrm{box})} \Pi^{w}(q) F^{v,w}(q) \ ,
  \label{e:Eq12}
\end{equation}
where $\delta_{v,w}$ is the Kronecker symbol (1 if $v = w$, otherwise zero), $\Pi^{w}(q)$ and $F^{v,w}(q)$ are the polarization factor and unit-less screening form factor, respectively \cite{esseni2011nanoscale}. In the case at study the dielectric matrix can be analytically inverted to evaluate screened matrix elements as:
\begin{equation}
M^{v}_\mathrm{scr}(q) =  \frac{\left(1 - \sum_{w\neq v} \epsilon^{v,w}(q) \right)M^{v}(q) + \sum_{w\neq v}\epsilon^{v,w}(q)M^{w}(q)}{2 - \sum_{w}\epsilon^{v,w}(q)} \ .
  \label{e:Eq13}
\end{equation}

The static dielectric function approach described above has been directly used for the scattering due to charged impurities, while the situation is admittedly more complicated for phonon scattering. For the inelastic, intervalley phonon transitions described in Table~\ref{t:DP} and Table~\ref{t:PH} the relatively large phonon wave-vector (see also \Fig{f:DP}) and the non-null phonon energies suggest that it is safe to leave these transitions unscreened, because the dynamic descreening and the large phonon wave-vectors make the screening very ineffective. Arguments concerning screening for intra-valley acoustic phonons are more subtle and controversial and a thorough discussion for inversion layer systems can be found in Ref.~\onlinecite{fischetti1993monte}. We here decided to leave also intra-valley acoustic phonons unscreened, which is the choice employed in essentially all the studies concerning transport in inversion layers that the authors are aware of. The screening of the SO phonon scattering is also a delicate  subject, because the polar phonon modes of the high-$\kappa$ dielectrics can couple with the collective excitations of the electrons in the MoS$_2$ layer and thus produce coupled phonon-plasmon modes \cite{ong2013theory,fischetti1993monte}, whose treatment is further complicated by the possible occurrence of Landau damping \cite{ong2013theory,fischetti1993monte,toniutti2012origin}. In this paper we do not attempt a full treatment of the coupled phonon-plasmon modes \cite{ong2013theory,fischetti1993monte}, but instead show in in~\Sec{s:RST} results for the two extreme cases of either unscreened SO phonons or SO phonons screened according to the static dielectric function. We can anticipate that while the inclusion of static screening in SO phonons implies a significant mobility enhancement compared to the unscreened case, the mobility dependence on the strain and on the dielectric constant of the high-$\kappa$ dielectrics is not significantly affected by the treatment of screening for SO phonons. 

\section{Mobility Calculation}
\label{s:MOB}
Acoustic, optical, polar-optical, remote phonon, and charged impurity scatterings are considered for the calculation of low-field mobility. As it will be discussed in the next section, the bandstructure for Q${_A}$ and Q${_B}$ valleys is not isotropic and the mobility shows direction-dependence, hence we calculated mobility by solving numerically the linearized Boltzmann Transport Equation (BTE) according to the approach described in Ref.~\onlinecite{paussa2013exact}, which does not introduce any simplifying assumption in the BTE solution. In particular, mobility has been calculated along the armchair and zigzag directions and strain has been also studied for the uniaxial configuration along either armchair or zigzag direction, as well as for the biaxial configuration. 
 
In order to describe in more detail the mobility calculation procedure, we first recall that the longitudinal direction of Q$_{A}$-valley is neither the armchair nor the zigzag direction, and \Fig{f:Ek0}(a) shows that $\theta$ is the angle describing the valley orientation with respect to the zigzag direction in k-space (i.l. armchair direction in real space). Let us now consider first the case of the mobility $\mu_{A}^{(v)}$ of valley $v$ along the armchair direction, that can be written by definition as $\mu_{A}^{(v)}  = J_{A}^{(v)}/F_{A}$, where $J_{A}^{(v)}$ is the current component in the armchair direction for the valley $v$ induced by the electric field $F_{A}$ along armchair direction. The current $J^{(v)}_{A}$ can be expressed as $J^{(v)}_{A} = J^{(v)}_{l}\cos{(\theta_{v})} + J^{(v)}_{t}\sin{(\theta_{v})}$ in terms of the current components $J^{(v)}_{l}$, $J^{(v)}_{t}$ along, respectively, the longitudinal and transverse direction of the valley $v$. By denoting the longitudinal ($F_{l}$) and transverse component ($F_{t}$) of the electric field as $F_{l} = F_{A}cos(\theta_{v})$ and $F_{t} = F_{A}\sin(\theta_{v})$, the currents $J_{l}^{(v)}$ and $J_{t}^{(v)}$ in turn can be written as $J^{(v)}_{l} = \mu^{(v)}_{ll}F_{l} + \mu^{(v)}_{lt}F_{t}$ and $J^{(v)}_t = \mu^{(v)}_{lt}F_{l} + \mu^{(v)}_{tt}F_{t}$, where $\mu_{ll}$, $\mu_{tt}$ and $\mu_{lt}$ are the entries of the two by two mobility matrix in the valley coordinate system \cite{esseni2011nanoscale}. Consequently we finally obtain:
\begin{equation}
 \mu^{(v)}_{A}  = \frac{J_{A}^{(v)}}{F_{A}} = \mu^{(v)}_{ll}\cos^2{(\theta_{v})} +  \mu^{(v)}_{tt}\sin^2{(\theta_{v})}  + 2\mu^{(v)}_{lt}\sin{(\theta_{v})}\cos{(\theta_{v})} \ .
  \label{e:MuA}
\end{equation}
By following a similar procedure, the mobility $\mu_{Z}^{(v)}$ of the valley $v$ along the zigzag direction can be written as:
\begin{equation}
 \mu^{(v)}_{Z}  = \frac{J_{A}^{(v)}}{F_{A}} = \mu^{(v)}_{ll}\sin^2{(\theta_{v})} +  \mu^{(v)}_{tt}\cos^2{(\theta_{v})}  - 2\mu^{(v)}_{lt}\sin{(\theta_{v})}\cos{(\theta_{v})} \ .
  \label{e:MuZ}
\end{equation}
For the circular and elliptical bands employed in our calculations (see \Fig{f:Ek}(g)), $\mu_{lt}^{(v)}$ is zero for symmetry reasons \cite{esseni2011nanoscale}. After calculating the mobility for each valley, the overall mobilities $\mu_{A}$ and $\mu_{Z}$ are obtained as the average of the mobility in the different valleys weighted by the the corresponding electron density. 

\Eq{e:MuA} and \Eq{e:MuZ} allow us to calculate the mobility $\mu_{A}^{(v)}$ and $\mu_{Z}^{(v)}$ from the longitudinal $\mu_{ll}^{v}$ and the transverse mobility $\mu_{tt}^{v}$ of the valley $v$ which are the mobilities obtained from the linearized BTE when the electric field is either in the longitudinal or in the transverse direction of the valley $v$. As already said, the $\mu_{ll}^{v}$ and $\mu_{tt}^{v}$ have been obtained by using the approach of Ref.~\onlinecite{paussa2013exact}, whose derivation for the case at study in this work can be summarized as follows. The out of equilibrium occupation function $f^{v}(\vec{k})$ for the valley $v$ in the presence of a field $F_{x}$ is written as
\begin{equation}
f^{v}(\vec{k}) = f_{0}(E^{v}(\vec{k})) - eF_{x}g^{v}(\vec{k}) \ ,
  \label{e:Eq14}
\end{equation}
where $f_{0}(E)$ is the equilibrium Fermi-Dirac distribution function, $x \in \{l,t \}$ is either the longitudinal or the transverse direction of the valley and $\vec{k} = (k_{l},k_{t})$ is the wavevector in the valley coordinate system. \Eq{e:Eq14} is a definition of $g^{v}(\vec{k})$, which is the unknown function of the linearized BTE problem. For a two-dimensional system the linearized BTE can be written as \cite{paussa2013exact}
\begin{equation}
\begin{split}
 g^{v}(\vec{k}) & \left( \frac{1}{2\pi \hbar} \sum_{w} \int_{\vec{k'}} \Lambda^{v,w}(\vec{k},\vec{k'})\delta[E^{w}(\vec{k'}) - E^{v}(\vec{k}) \mp \hbar \omega_{ph}]\vec{dk'} \right) \\
 & -\frac{1}{2 \pi \hbar} \sum_{w} \int_{\vec{k'}} \Lambda^{v,w}(\vec{k},\vec{k'}) g^{w}(\vec{k'}) \delta[E^{w}(\vec{k'}) - E^{v}(\vec{k}) \mp \hbar \omega] \vec{dk'} = v^{v}_{x}(\vec{k}) \ ,
\end{split}
  \label{e:Eq151}
\end{equation}
where $v^{v}_{x}$ is the $x$ component of the group velocity of valley $v$ and, for convenience of notation, we have introduced the quantity
\begin{equation}
\Lambda^{v,w}(\vec{k},\vec{k'}) = |M^{v,w}(\vec{k},\vec{k'})|^{2} \left[\frac{1-f_{0}(E^{w}(\vec{k'}))}{1-f_{0}(E^{v}(\vec{k}))}  \right] \ .
  \label{e:Eq152}
\end{equation}
To numerically solve \Eq{e:Eq151}, we employed the discretization scheme introduced in Ref.~\onlinecite{paussa2013exact}: $\vec{k}$ is discretized according to a uniform angular step $\Delta \beta$ and also a uniform energy step. The discrete values $k_{v,r,d}$ of the wave-vector magnitude correspond to one of the discrete energy values and the generic discrete wave-vector $\vec{k}_{v,r,d} = (k_{v,r,d},d\Delta\beta)$ is identified by the magnitude $k_{v,r,d}$ and the angle $d\Delta\beta$ (with $d$ being a positive integer number). For each scattering mechanism, by converting the integral over $\vec{k}$ in an integral over the energy and the angle $\beta$ and then using the above mentioned discretization, Eq.~\ref{e:Eq151} can be rewritten as:
\begin{equation}
\begin{split}
 &g(k_{v,r,d}) \left[\frac{\Delta\beta}{2\pi\hbar} \sum_{w,r',d'}A_{v,r,d}^{w,r',d'}\delta_{v,r,d}^{w,r',d'}\right] - \frac{\Delta\beta}{2\pi\hbar} \sum_{w,r',d'}B_{v,r,d}^{w,r',d'} g(k_{w,r',d'}) \delta_{v,r,d}^{w,r',d'}  \\ 
&= \frac{v_{x}(k_{v,r,d})f(E(k_{v,r,d}))[1-f(E(k_{v,r,d}))]}{k_\mathrm{B}T} \ .
\end{split}
\label{e:Eq18}
\end{equation}
Eq.~\ref{e:Eq18} is a linear problem for the discretized unknown values $g(k_{v,r,d})$ written in terms of the coefficients $A_{v,r,d}^{w,r',d'}$ and $B_{v,r,d}^{w,r',d'}$ defined as
\begin{equation}
A_{v,r,d}^{w,r',d'} = k_{w,r',d'} \left[\frac{dE(k_{w,r',d'})}{dk} \right]^{-1}  \times |M^{v,w}(k_{v,r,d},k_{w,r',d'})|^{2} \left[\frac{1-f_{0}(E(k_{w,r',d'}))}{1-f_{0}(E(k_{v,r,d}))}\right] \ ,
  \label{e:Eq19}
\end{equation}
\begin{equation}
B_{v,r,d}^{w,r',d'} = k_{w,r',d'}\left[\frac{dE(k_{w,r',d'})}{dk} \right]^{-1} \times |M^{v,w}(k_{v,r,d},k_{w,r',d'})|^{2} \left[\frac{f_{0}(E(k_{v,r,d}))}{f_{0}(E(k_{w,r',d'}))} \right] \ ,
  \label{e:Eq20}
\end{equation}
where the non-zero entries of the matrix representing the linear problem are governed by the Kronecker symbols $\delta_{v,r,d}^{w,r',d'}$, that are defined so to enforce energy conservation \cite{paussa2013exact}.

Eq.~\ref{e:Eq18} has been written for a single scattering mechanism. In order to accommodate several scattering mechanisms in our calculations, we do not resort to an approximated treatment based on the Matthiessen rule \cite{esseni2011quantitative}, but instead follow Ref.~\onlinecite{paussa2013exact} and notice that Eq.~\ref{e:Eq18} can be written in the concise matrix notation $\bar{\bar{M}}^{(s)} \bar{g} = \bar{G}$, where $\bar{\bar{M}}^{(s)}$ is a matrix specific of the scattering mechanisms $s$, $\bar{g}$ is the unknown vector and $\bar{G}$ is the vector at the right hand side of Eq.~\ref{e:Eq18} and consisting of known quantities. Hence the unknown vector $\bar{g}$ corresponding to several scattering mechanisms can be obtained by solving the linear problem 
\begin{equation}
\left[\sum_{s=1}^{N_\mathrm{SC}} \bar{\bar{M}}^{(s)} \right] \bar{g} = \bar{G} \ ,
\label{e:Eq17}
\end{equation}
where Eq.~\ref{e:Eq18}--Eq.~\ref{e:Eq20} will totally define the entries of the matrix $\bar{\bar{M}}^{(s)}$ for each scattering mechanism.

%
\section{Results and Discussions}
\label{s:RST}
\Fig{f:Ek} shows the energy contours of the conduction-band in the first BZ for strained single layer MoS$_{2}$. In an unstrained material the lowest and the second lowest minimum in the conduction band are denoted as K-valley and Q-valley, respectively. The 6 Q-valleys are degenerate for unstrained and biaxial strain conditions. With the application of uniaxial strain, however, they split into 4 Q$_\mathrm{A}$-valleys and 2 Q$_\mathrm{B}$-valleys with different effective masses and energy minima. \Fig{f:Band} illustrates the bandstructure of unstrained and strained single layer MoS$_{2}$ including K, Q$_\mathrm{A}$, and Q$_\mathrm{B}$ valleys. Under compressive strain one of the Q$_\mathrm{A}$ or Q$_\mathrm{B}$ valleys becomes the lowest valley.

The energy distance between these K-valley and Q-valley for unstrained material is evaluated to be 160 meV, in agreement with Ref. \onlinecite{kaasbjerg2012phonon}. Tensile strain increases this energy distance, which is instead reduced by a compressive strain. In particularly, a relatively large compressive strain lowers the energy of Q-valley so that it becomes the lowest valley as shown in \Fig{f:Efm}(a)-(c). Here we can anticipate that, while under tensile strain one can neglect the scattering between Q and K-valleys, under compressive strain this type of scattering can significantly affect the mobility. Assuming a non-parabolic dispersion relation $E(1+ \alpha E )=\hbar^{2}k_{l}^{2}/2m_{l}^{*} + \hbar^{2}k_{t}^{2}/2m_{t}^{*}$, the longitudinal $m_{l}^{*}$ and transverse $m_{t}^{*}$ effective mass and also the non-parabolicity factor $\alpha$ are extracted from the DFT-calculated electronic bandstructure and reported in \Fig{f:Efm}(d)-(f). As can be seen in \Fig{f:Efm}(a)-(c), under compressive uniaxial strain the energy minima of all K- and Q-valleys are quite close, while at large compressive biaxial strain the K-valley lie at higher energy and their contribution to mobility can be neglected.

We compare in Table~\ref{t:Comp} our calculated mobilities at various carrier concentrations with the experimental data reported in Ref.~\onlinecite{radisavljevic2013mobility} for unstrained single-layer MoS$_2$ embedded between SiO$_{2}$ and HfO$_{2}$ with impirity density 4 $\times 10^{12}$. At $T=100$ K the effect of piezoelectric can be ignored \cite{kaasbjerg2013acoustic}. Very good agreement with experimental data validates the  bandstructure and mobility models employed in this work.

The strain-dependency of intrinsic phonon limited mobility is presented in \Fig{f:Mob}(a). Apparently, the effects of compressive and tensile strain on mobility are very different, which can be mainly explained by considering the role of inter-valley scattering. For example, with tensile strain the minimum energies of Q$_{A}$ and Q$_{B}$-valleys are much higher than that of K-valley, which suppresses inter-valley scattering. Under compressive strain, instead, the inter-valley scattering cannot be neglected because of the smaller energy difference between these valleys. With tensile biaxial strain, the mobility increases because of the reduction of the effective mass and also the increase of the energy difference between K and Q-valleys, which results in the reduction of the inter-valley scattering rate. With a tensile biaxial strain of $5\%$ the phonon limited mobility becomes $75\%$ higher than that of unstrained material. In contrast, a compressive biaxial strain of 0.8$\%$ strongly reduces the mobility due to the reduction of energy difference between K and Q-valleys (see \Fig{f:Efm}(a)) and increased inter-valley scattering. With further increase of compressive biaxial strain, Q-valleys become the lowest ones and thus dominate the mobility. At a strain value of about $2.5\%$ the contribution of K-valleys to mobility becomes negligible and the mobility behavior is completely determined by the Q-valleys. Longitudinal and transverse effective masses of Q-valleys are not equal and are somewhat changed by strain, however, the different angular dependency of mobility along the armchair and zigzag direction tends to compensate the changes of effective masses and the overall mobility remains nearly constant at larger compressive strain values.

Under tensile uniaxial strain the mobility is hardly affected by a strain along the zigzag direction, while it increases for strain along the armchair direction. In both cases the variation of the effective mass and non-parabolicity factor with strain determine the mobility behavior. Under a compressive uniaxial strain along the armchair direction, Q$_\mathrm{A}$ becomes the lowest valley, while for a strain along the zigzag direction Q$_\mathrm{B}$ is the lowest one. These results emphasize that the contribution of both Q$_\mathrm{A}$ and Q$_\mathrm{B}$ valley should be included for an accurate calculation of mobility. Under a compressive strain of about 1.5\% the mobilities are strongly reduced, but they remain nearly constant for larger strain magnitudes. Moreover, we notice that for a strain along the zigzag direction, the mobility along the strain direction becomes slightly larger than the mobility in the armchair direction.

\Fig{f:Mob}(b) reports the mobility in the presence of intrinsic phonon and charged impurity scattering. The top and bottom oxide are assumed to be SiO$_2$ and both carrier and impurity concentrations are $10^{12}$ cm$^{-2}$. Except for a global reduction of the mobility, the behavior of the mobility with strain is similar to \Fig{f:Mob}(a) corresponding to phonon limited mobility. The results presented in \Fig{f:Mob}(c) correspond to the same parameters as in \Fig{f:Mob}(b), except for a reduction of carrier concentration to $10^{11}$  cm$^{-2}$. As the carrier concentration decreases the effect of static screening becomes weaker and the mobility is further reduced. \Fig{f:Mob}(d) illustrates the mobility as a function of strain with the same parameters used in \Fig{f:Mob}(b), expect for the top and bottom gate oxide which is Al$_2$O$_3$. A high-$\kappa$ dielectric implies a larger dielectric screening and increases the mobility. Under this condition, with a tensile biaxial strain of 5\% and a tensile uniaxial strain of 5\% along the armchair direction the mobility increases by 53\% and 43\%, respectively, compared to an unstrained single-layer MoS$_2$. For a better comparison, \Fig{f:Comp} shows the room temperature mobility versus carrier concentration and also versus the dielectric constant for the unstrained material and for 5\% tensile strain in either a biaxial or a uniaxial configuration along the armchair direction with an impurity  density equal to $10^{12}$ cm$^{-2}$. As can be seen in \Fig{f:Comp}(a), because of screening the mobility increases with the carrier concentration for both unstrained and strained cases. \Fig{f:Comp}(b) indicates that the strain induced mobility enhancement with high-$\kappa$ dielectric materials is slightly larger than that with low-$\kappa$ materials.

The effect of unscreened and screened remote phonon scattering on the mobility of unstrained and 5\% biaxial strained single layer MoS$_{2}$ are compared in \Fig{f:Remote}. Except for a global increase of mobility values. The mobility dependence on the dielectric constant $\kappa$ is not significantly affected by the screening of SO phonons. As can be seen, for relatively small $\kappa$ values, mobility improves with increasing $\kappa$ because of the dielectric screening of charged impurities \cite{ma2014charge}. At high $\kappa$ values, however, the mobility decrease with increasing $\kappa$ because the corresponding smaller SO phonon energies (see Table~\ref{t:Remote}) tend to increase momentum relaxation time via SO phonons. For the conditions considered in \Fig{f:Remote}(temperature, carrier and impurity concentrations, and semi-infinite dielectrics with SiO$_2$ as the bottom oxide), AlN appears to be the optimal top dielectric material for strained and also unstrained single layer MoS$_{2}$. \Fig{f:Temp} shows the temperature dependency of the mobility for unstrained and 5\% biaxial strain with HfO$_{2}$ as the top oxide. As expected the effect of inelastic remote phonons increases with temperature for both unstrained and strained cases. Therefore, it is expected that the optimal material as a top dielectric for temperatures above(bellow) 300 K, should have a lower(higher)-$\kappa$ compared to AlN.
  
\section{Conclusion}
\label{s:CON}
A comprehensive theoretical study on the role of strain on the mobility of single-layer MoS$_2$ is presented. DFT calculations are used to obtain the effective masses and energy minima of the contributing valleys. Thereafter, the linearized BTE is solved for evaluating the mobility, including the effect of intrinsic phonons, remote phonons, and screened charged impurities. The results indicate that, a tensile strain increases the mobility, while compressive strain reduces the mobility. Furthermore, biaxial strain and uniaxial strain along the armchair direction increase the mobility more effectively. The strain-dependency of the mobility of MoS$_2$ is rather complicated and strongly depends on the relative positions of Q and K-valleys and the corresponding inter-valley scattering.  The presented results pave the way for a possible strain engineering of the electronic transport in MoS$_2$ based electron devices.

\newpage
%

\newpage
\clearpage

\begin{table}[!h]
\caption{Deformation potentials for inelastic phonon assisted transitions in single layer MoS$_{2}$. All parameters are taken from Ref.~\onlinecite{li2013intrinsic}.}
\begin{ruledtabular}
\begin{tabular}{c c c }
Phonon momentum & Electron transition & Deformation potential
\\ \hline
$\Gamma$ & K$ \rightarrow $K & $D_\mathrm{ac} = 4.5$ eV \\
$\Gamma$ & K$ \rightarrow $K & $D_\mathrm{op} = 5.8 \times 10^{8}$ eV/cm \\

K & K$ \rightarrow $K$'$ & $D_\mathrm{ac} = 1.4 \times 10^{8}$ eV/cm \\
K & K$ \rightarrow $K$'$ & $D_\mathrm{op} = 2.0 \times 10^{8}$ eV/cm \\

Q & K$ \rightarrow $Q & $D_\mathrm{ac} = 9.3 \times 10^{7}$ eV/cm \\
Q & K$ \rightarrow $Q & $D_\mathrm{op} = 1.9 \times 10^{8}$ eV/cm \\

M & K$ \rightarrow$ Q & $D_\mathrm{ac} = 4.4 \times 10^{7}$ eV/cm \\
M & K$ \rightarrow$ Q & $D_\mathrm{op} = 5.6 \times 10^{8}$ eV/cm \\
\\ \hline

$\Gamma$ & Q $\rightarrow$ Q & $D_\mathrm{ac} = 2.8$ eV \\
$\Gamma$ & Q $\rightarrow$ Q & $D_\mathrm{op} = 7.1 \times 10^{8}$ eV/cm \\

Q & Q $ \rightarrow $Q & $D_\mathrm{ac} = 2.1 \times 10^{8}$ eV/cm \\
Q & Q $ \rightarrow $Q & $D_\mathrm{op} = 4.8 \times 10^{8}$ eV/cm \\

M & Q $\rightarrow $Q & $D_\mathrm{ac} = 2.0 \times 10^{8}$ eV/cm \\
M & Q $\rightarrow $Q & $D_\mathrm{op} = 4.0 \times 10^{8}$ eV/cm \\

K & Q $\rightarrow$ Q & $D_\mathrm{ac} = 4.8 \times 10^{8}$ eV/cm \\
K & Q $\rightarrow$ Q & $D_\mathrm{op} = 6.5 \times 10^{8}$ eV/cm \\

Q & Q $\rightarrow $K or K$'$ & $D_\mathrm{ac} = 1.5 \times 10^{8}$ eV/cm \\
Q & Q $\rightarrow $K or K$'$ & $D_\mathrm{op} = 2.4 \times 10^{8}$ eV/cm \\

M & Q $\rightarrow $K or K$'$ & $D_\mathrm{ac} = 4.4 \times 10^{8}$ eV/cm \\
M & Q $\rightarrow $K or K$'$ & $D_\mathrm{op} = 6.6 \times 10^{8}$ eV/cm \\
\end{tabular}
\end{ruledtabular}
\label{t:DP}
\end{table}

\newpage
\begin{table}[!h]
\caption{Phonon energy for intra-valley and inter-valley transitions at the K, M, and Q points of single layer MoS$_{2}$ as reported in Ref.~\onlinecite{li2013intrinsic}. As discussed in Ref.~\onlinecite{li2013intrinsic}, the energy values for acoustic (optical) phonon modes is the average of phonon energies of transverse and longitudinal (transverse, longitudinal and homo-polar) modes. }
\begin{ruledtabular}
\begin{tabular}{c c c c c}
Phonon mode & $\Gamma$ & K & M & Q
\\ \hline
Acoustic phonon energy [meV] & 0 & 26.1 & 24.2 & 20.7\\
Optical phonon energy [meV] & 49.5 & 46.8 & 47.5 & 48.1  \\
\end{tabular}
\end{ruledtabular}
\label{t:PH}
\end{table}

\newpage
\begin{table}[!h]
\caption{Parameters for the dielectric materials taken from (a) Ref.~\onlinecite{konar2010effect} and (b) Ref.~\onlinecite{perebeinos2010inelastic} and corresponding calculated SO phonon frequencies $\hbar \omega_\mathrm{so,tox}$ and $\hbar \omega_\mathrm{so,box}$. In all of the cases an SiO$_{2}$ bottom oxide is assumed.}
\begin{ruledtabular}
\begin{tabular}{c c c c c c c}
Top oxide dielectric material & SiO$_{2}^{(a)}$ & BN$^{(b)}$ & AlN$^{(a)}$ & Al$_{2}$O$_{3}^{(a)}$ & HfO$_{2}^{(a)}$ & ZrO$_{2}^{(a)}$
\\ \hline
$\epsilon^{0}_\mathrm{tox}$ & 3.9 & 5.09 & 9.14 & 12.53 & 23 & 24 \\
$\epsilon^{\infty}_\mathrm{tox}$ & 2.5 & 4.1 & 4.8 & 3.2 & 5.03 & 4 \\
$\omega_\mathrm{TO,tox}$ [meV]& 55.6 & 93.07 & 81.4 & 48.18 & 12.4 & 16.67 \\
$\omega_\mathrm{so,tox}$ [meV](Evaluated in this work) & 69.4 & 100.5 & 104.3 & 83.9 & 21.3 & 30.5 \\
$\omega_\mathrm{so,box}$ [meV](Evaluated in this work) & 69.4 & 60.1 & 58.0 & 54.2 & 61.1 & 62.9 \\
\end{tabular}
\end{ruledtabular}
\label{t:Remote}
\end{table}

\newpage
\begin{table}[!h]
\caption{Comparison of the calculated mobility in this work with the experimental data of Ref.~\onlinecite{radisavljevic2013mobility}. $T=100$ K and the impurity density is $4\times10^{12}$ cm$^{-2}$.}
\begin{ruledtabular}
\begin{tabular}{c c c c c}
Carrier concentration [cm$^{-2}$] & $7.6\times10^{12}$ & $9.6\times10^{12}$ & $1.15\times10^{13}$ & $1.35\times10^{13}$
\\ \hline
Calculated mobility, this work [cm$^2$/(Vs)] & 93 & 106  & 114  & 122\\
Experimental mobility [cm$^2$/(Vs)] & 96$\pm$3 & 111$\pm$3 & 128$\pm$3  & 132$\pm$3
\end{tabular}
\end{ruledtabular}
\label{t:Comp}
\end{table}
\newpage
\clearpage
\section*{List of Figures}	
\begin{figure}[!ht]
  \begin{center}
    \includegraphics[width=8cm]{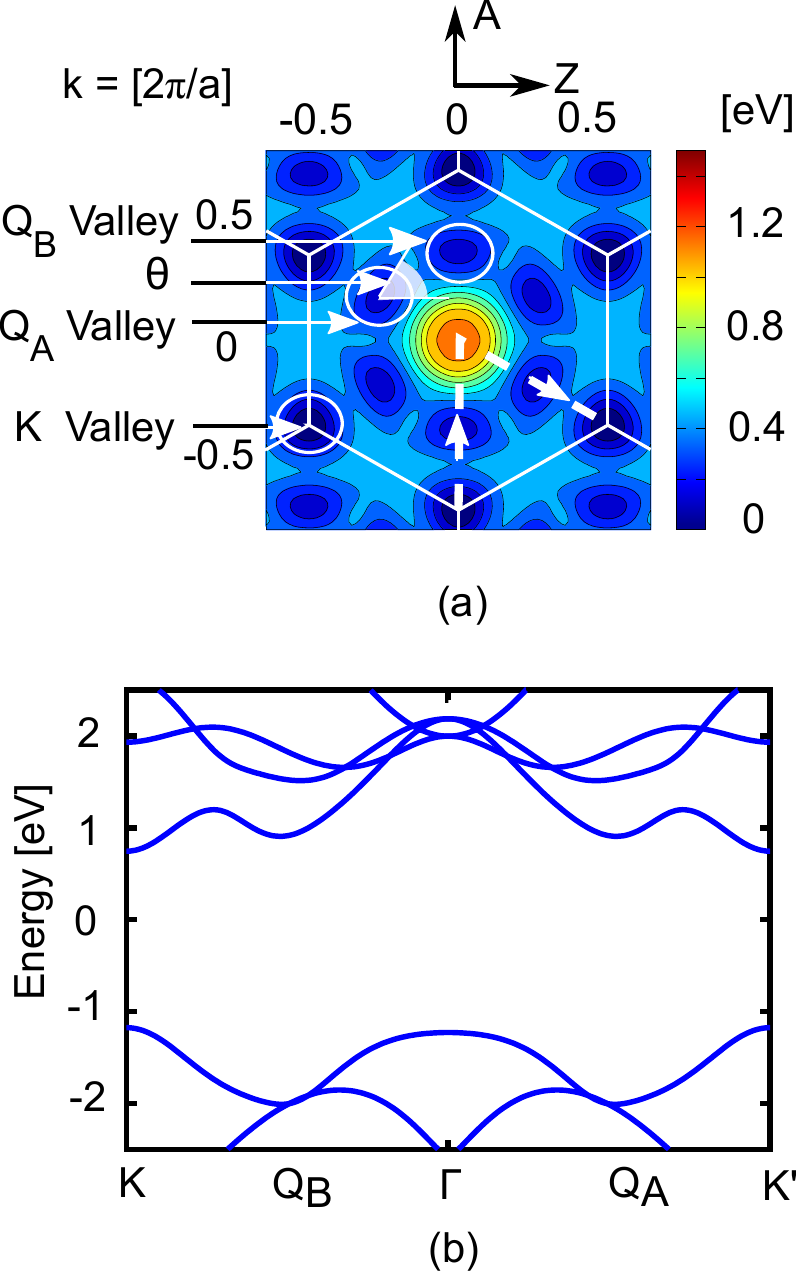}
  \end{center}
  \caption{\label{f:Ek0}}
\end{figure}

\begin{figure}[!ht]
  \begin{center}
    \includegraphics[width=10cm]{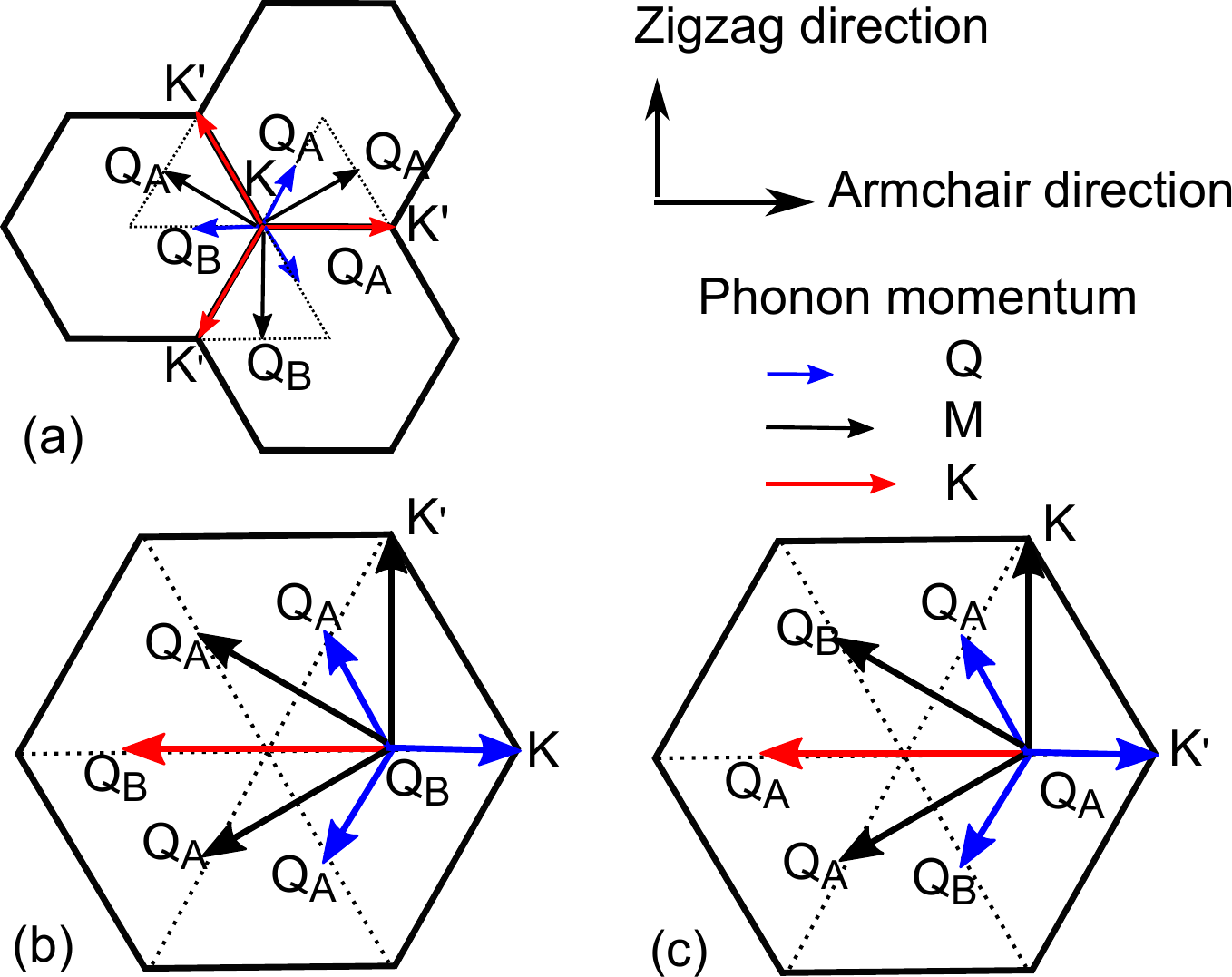}
  \end{center}
  \caption{\label{f:DP}}
\end{figure}

\begin{figure}[!ht]
  \begin{center}
    \includegraphics[width=8cm]{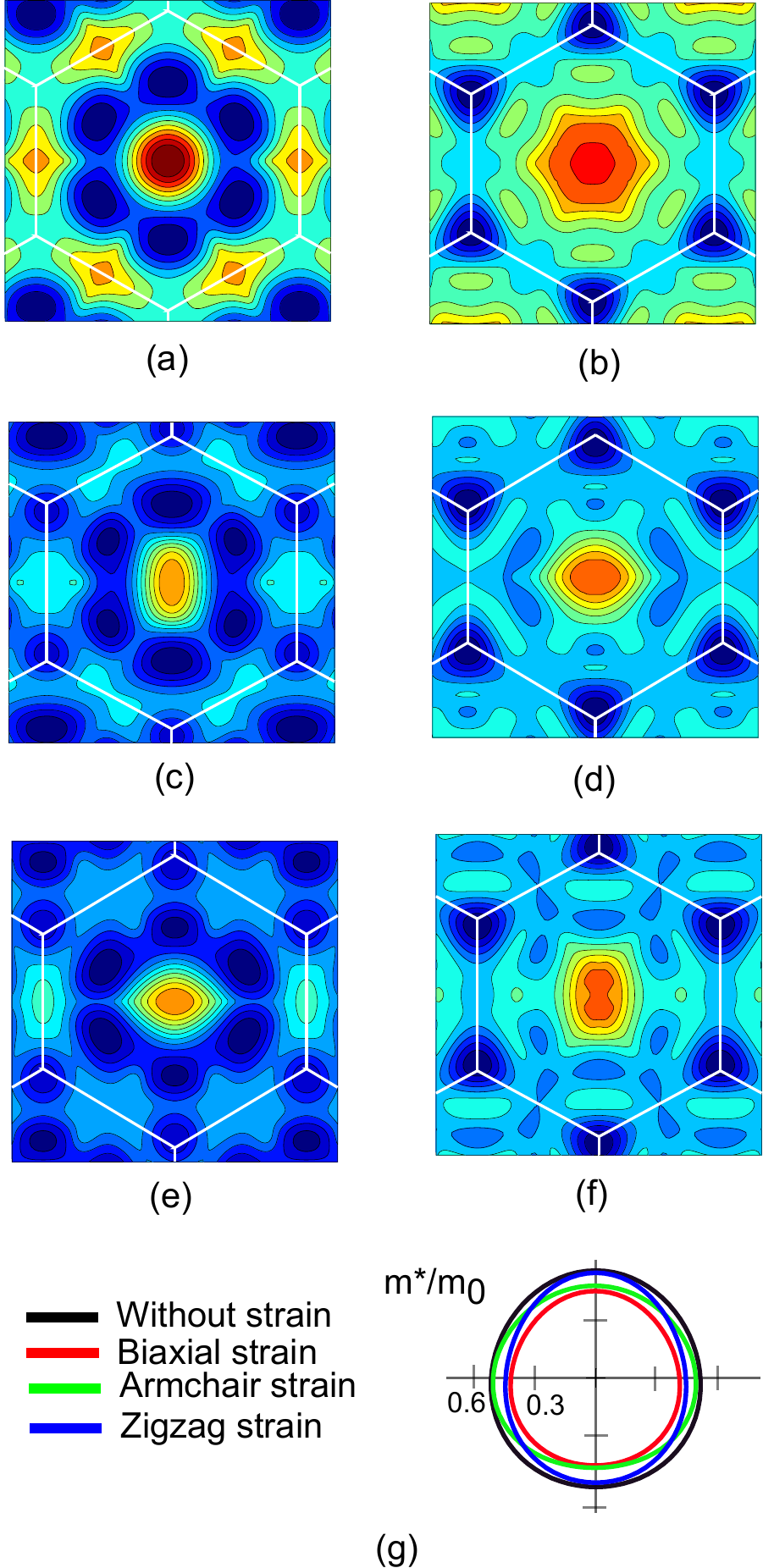}
  \end{center}
  \caption{\label{f:Ek}}
\end{figure}

\begin{figure}[!ht]
  \begin{center}
    \includegraphics[width=8cm]{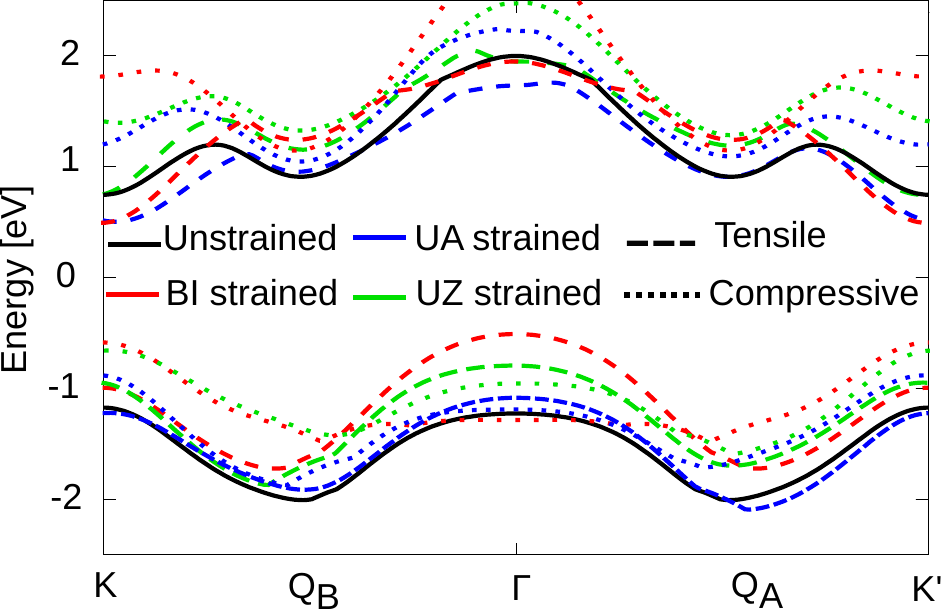}
  \end{center}
  \caption{\label{f:Band}}
\end{figure}

\begin{figure}[!ht]
  \begin{center}
    \includegraphics[width=14cm]{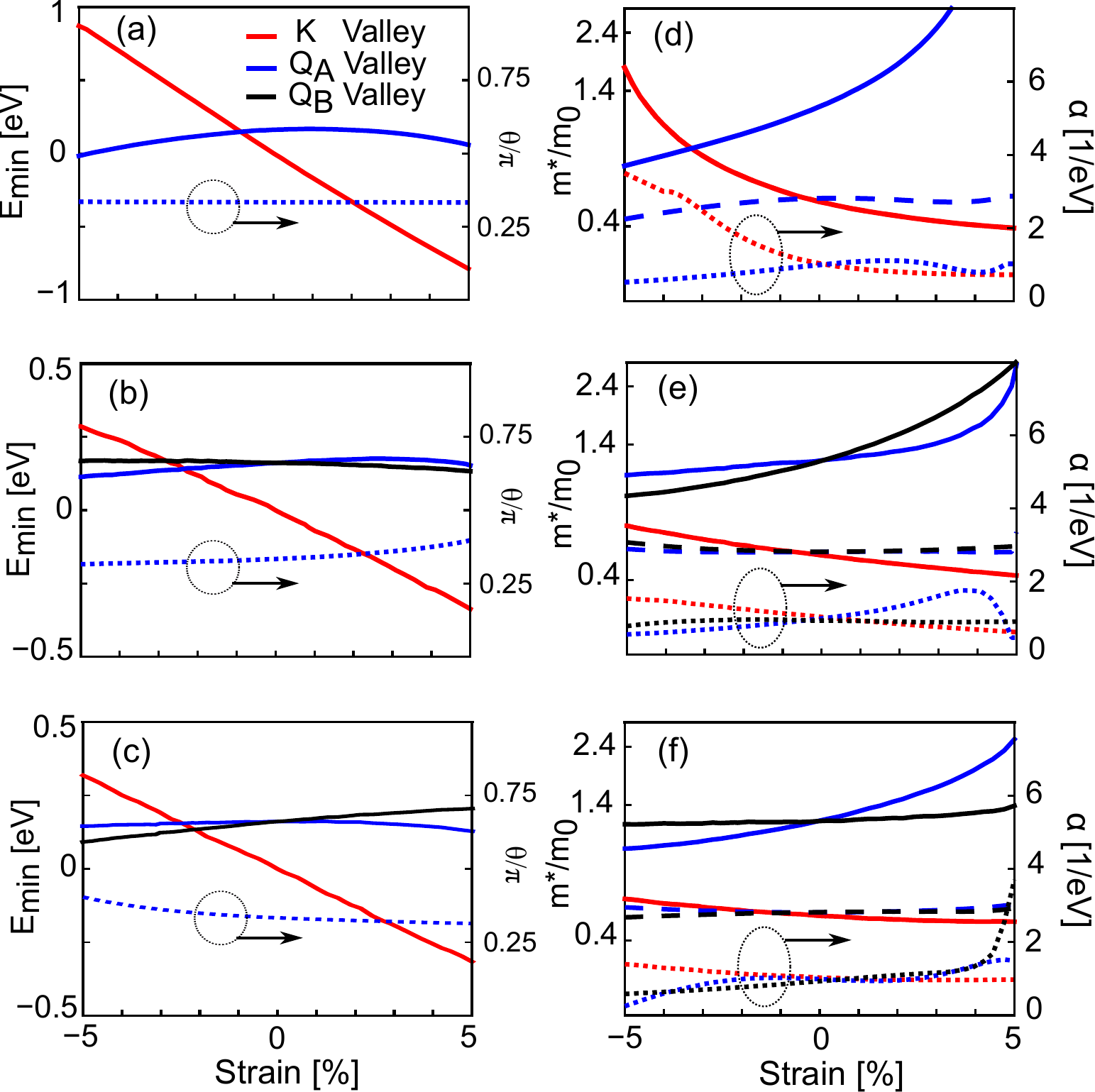}
  \end{center}
  \caption{\label{f:Efm}}
\end{figure}

\begin{figure}[!ht]
  \begin{center}
    \includegraphics[width=14cm]{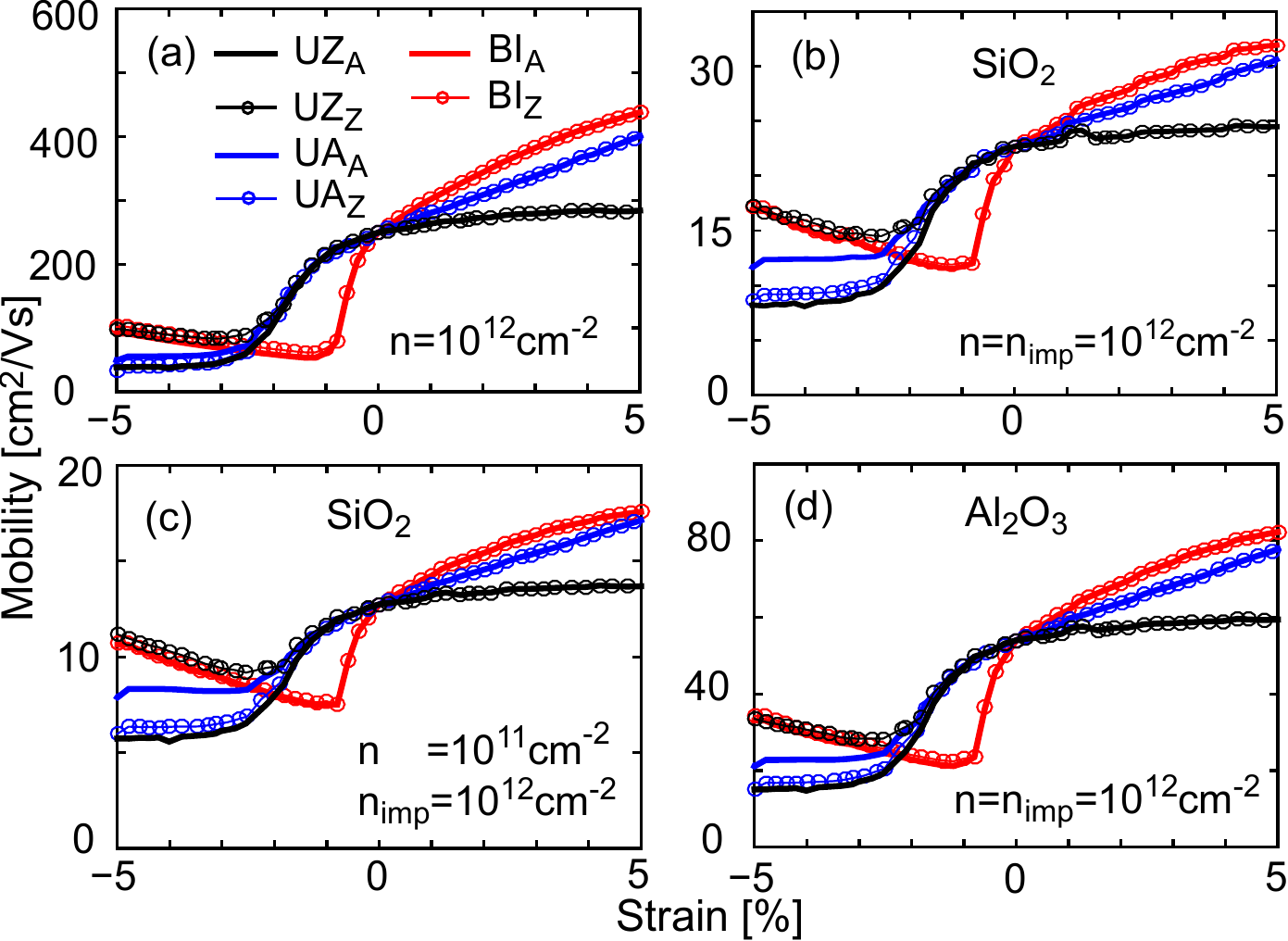}
  \end{center}
  \caption{\label{f:Mob}}
\end{figure}

\begin{figure}[!ht]
  \begin{center}
    \includegraphics[width=14cm]{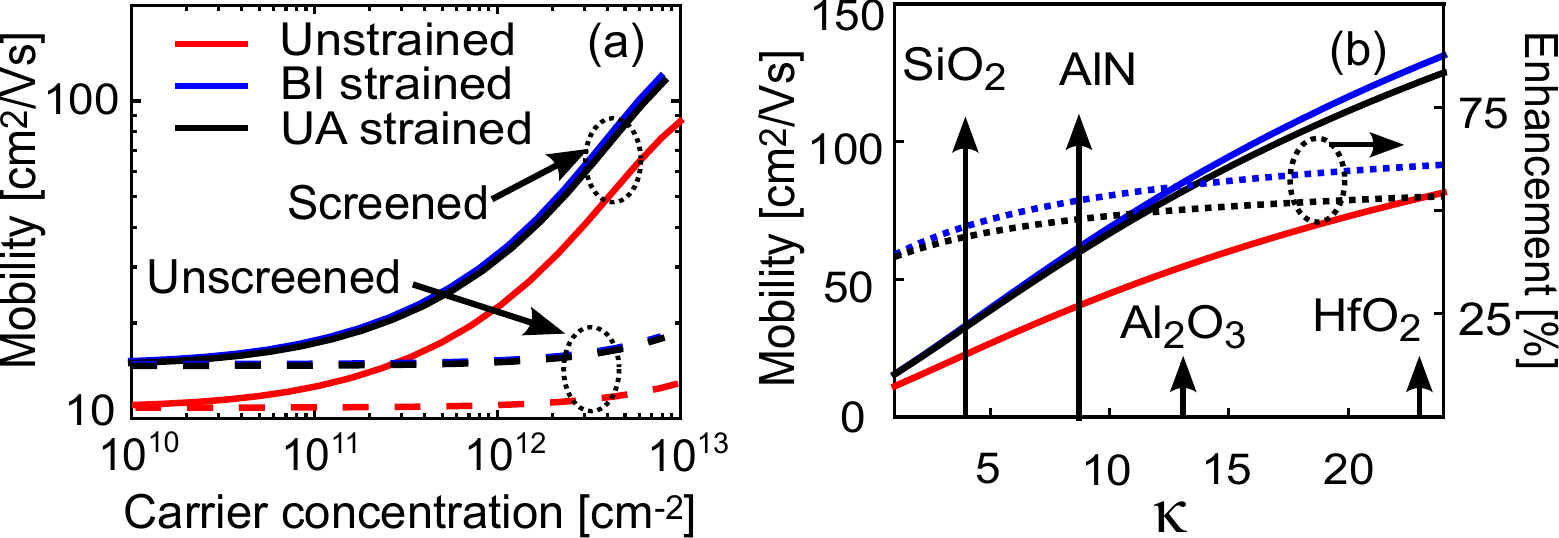}
  \end{center}
  \caption{\label{f:Comp}}
\end{figure}

\begin{figure}[!ht]
  \begin{center}
    \includegraphics[width=8cm]{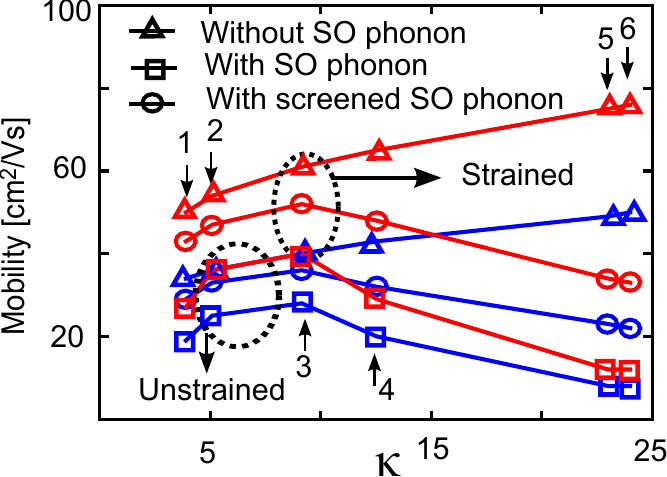}
  \end{center}
  \caption{\label{f:Remote}}
\end{figure}

\begin{figure}[!ht]
  \begin{center}
    \includegraphics[width=8cm]{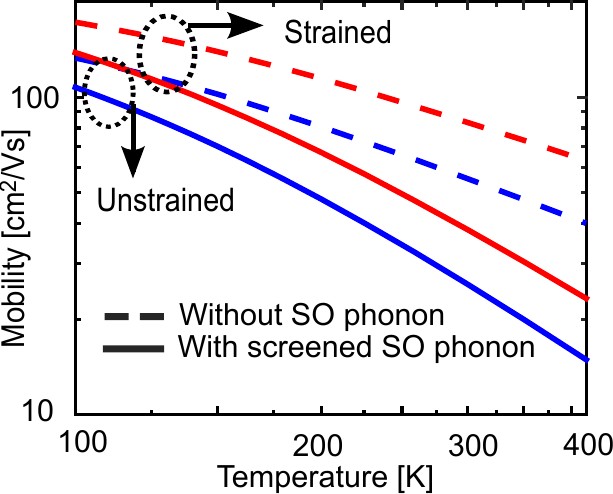}
  \end{center}
  \caption{\label{f:Temp}}
\end{figure}

\begin{description}
\item[\Fig{f:Ek0} ]{(a) Equi-energy contours in the first Brillouin zone for the unstrained single layer MoS$_{2}$. The angle $\theta$ that describes the Q$_\mathrm{A}$ valleys orientation in $\vec{k}$-space is also depicted in the figure. It should be recalled that the zigzag direction in $\vec{k}$-space corresponds to the armchair direction in real space. (b) The bandstructure of unstrained single layer MoS$_{2}$ in the first Brillouin zone and along the symmetry directions that are illustrated in (a).}

\item[\Fig{f:DP} ]{Illustration of several phonon assisted inter-valley transitions in single layer MoS$_{2}$ for (a) transitions from K-valley to other valleys; (b) transitions from Q$_{A}$-valley to other valleys; (c) transitions from Q$_{B}$-valley to other valleys. The figure also sets the notation used in Table~\ref{t:DP} and Table~\ref{t:PH} to identify phonon assisted transitions.}

\item[\Fig{f:Ek} ]{Equi-energy contours for single layer MoS$_{2}$ under: (a) compressive biaxial strain; (b) tensile biaxial strain; (c) compressive uniaxial strain along the armchair direction;
(d) tensile uniaxial strain along the armchair direction; (e) compressive uniaxial strain along the zigzag direction; (f) tensile uniaxial strain along the zigzag direction. (g) Extracted effective mass of K-valley along all directions in polar coordinate for unstrained MoS$_{2}$ and under tensile biaxial and uniaxial strain along armchair and zigzag directions. The nearly circular shape of the effective mass plot justifies the assumption of isotropic bandstructure. The strain magnitude is 4\% in all strained cases. The longitudinal and transverse effective masses of Q-valleys vary with the strain conditions.}

\item[\Fig{f:Band} ]{The band structure of unstrained and strained single layer MoS$_{2}$. BI: biaxial strain, UA: uniaxial strain along armchair direction; UZ: uniaxial strain along zigzag direction. The strain magnitude is 4\% in all strained cases.}

\item[\Fig{f:Efm} ]{ The minimum energies of valleys (solid-lines) and the angle $\theta$ (dotted lines) between the longitudinal direction of Q$_{A}$ valleys and zigzag direction in $k$-space as illustrated in \Fig{f:Ek0}(a) under: (a) biaxial strain; (b) uniaxial strain along the armchair direction; (c) uniaxial strain along the zigzag direction. The $\theta$ angle in \Fig{f:Efm}(a)-(c) corresponds to the Q$_{A}$ valley indicated in \Fig{f:Ek0}(a), and the $\theta$ angle of the other Q$_{A}$ valleys can be inferred from symmetry considerations. The $\theta$ angle for Q$_{B}$ valleys has a negligible dependence on strain (not shown) and it is approximately zero (see \Fig{f:Ek0}(a)). The effective masses (solid-lines for longitudinal and dashed-lines for transverse) and the non-parabolicity factor ($\alpha$) (dotted-lines) of various valleys under: (d) biaxial strain; (e) uniaxial strain along the armchair direction; (f) uniaxial strain along the zigzag direction. The longitudinal and transverse effective masses of K-valley are assumed to be equal.}

\item[\Fig{f:Mob}]{(a) Phonon limited mobility  of single layer MoS$_{2}$ as a function of strain with a carrier concentration $n = 10^{12}$ cm$^{-2}$. Mobility limited by phonon and screened charged impurity scattering with SiO$_2$ as the gate oxide ($\epsilon_\mathrm{r}$ = 3.9) and carrier ($n$) and charged impurity concentration ($n_\mathrm{imp}$) for: (b) $n=n_\mathrm{imp}=10^{12}$ cm$^{-2}$; (c) $n= 10^{11}$ cm$^{-2}$ and $n_\mathrm{imp} = 10^{12}$ cm$^{-2}$. (d) Same as (b), except for the gate oxide which is Al$_2$O$_3$. In the legend, BI, UA, and UZ denote biaxial strain, uniaxial strain along the armchair direction, and uniaxial strain along the zigzag direction respectively. The subscripts $A$ and $Z$ indicate the component of the mobility along the armchair or zigzag direction. For example: UZ$_{A}$ is the mobility along armchair direction for a uniaxial strain along zigzag direction.}

\item[\Fig{f:Comp}  ]{(a) The mobility versus carrier concentration with and without screening for the unstrained MoS$_{2}$, for a tensile biaxial strain of 5\%, and for a uniaxial strain of 5\% along the armchair direction. $n_\mathrm{imp} = 10^{12}$ cm$^{-2}$. (b) The mobility versus the relative dielectric constant for unstrained MoS$_{2}$ and for strain conditions as in (a). $n=n_\mathrm{imp} = 10^{12}$ cm$^{-2}$. The strain induced mobility enhancement is shown on the right-side of the $y$-axis.}

\item[\Fig{f:Remote} ]{The mobility accounting  for intrinsic phonon and charged impurity scattering (triangle), and for either unscreened (rectangle) or screened (circle) SO phonon scattering as a function of top oxide dielectric constant for unstrained (blue line) and 5\% biaxial strain (red line). Numbers 1 to 6 indicate the $\kappa$ value corresponding to dielectric materials studied in this work (see also Table~\ref{t:Remote}). In particular, (1): SiO$_{2}$, (2): BN, (3): AlN, (4): Al$_{2}$O$_{3}$, (5): HfO$_{2}$, and (6): ZrO$_{2}$. In all cases the back oxide is assumed to be SiO$_{2}$. T = 300 K, the impurity and carrier concentrations are equal to $4 \times 10^{12}$ cm$^{-2}$ and $10^{13}$ cm$^{-2}$, respectively. These values are consistent with experimental data reported in Ref.~\onlinecite{radisavljevic2013mobility}.} 

\item[\Fig{f:Temp} ]{The mobility with the inclusion of intrinsic phonon and charged impurity scattering (dash line) and with the inclusion of screened SO phonon (solid line) versus temperature for a SiO$_{2}/$MoS$_{2}/$HfO$_{2}$ structure for unstrained (blue line) and 5\% biaxial strain (red line). The impurity and carrier concentrations are equal to $4 \times 10^{12}$ cm$^{-2}$ and $10^{13}$ cm$^{-2}$, respectively.} 

\end{description}

\clearpage
\newpage
\vspace*{2cm}

\clearpage
\newpage
\vspace*{2cm}

\clearpage
\newpage
\vspace*{2cm}

\clearpage
\newpage
\vspace*{2cm}

\clearpage
\newpage
\vspace*{2cm}

\clearpage
\newpage
\vspace*{2cm}

\clearpage
\newpage
\vspace*{2cm}

\clearpage
\newpage
\vspace*{2cm}

\clearpage
\newpage
\vspace*{2cm}

\end{document}